\documentclass[aps, pra, 10pt, twocolumn, superscriptaddress, floatfix,longbibliography]{revtex4-1}
\usepackage[english]{babel}

\usepackage{array}

\newcolumntype{C}{>{$\displaystyle}c<{$}}
\newcolumntype{L}{>{$\displaystyle}l<{$}}
\newcolumntype{R}{>{$\displaystyle}r<{$}}
\usepackage{multirow}

\usepackage[plainpages=false,colorlinks=true,linkcolor=blue, citecolor=blue, urlcolor=blue]{hyperref}
\usepackage[charter]{mathdesign}
\usepackage{amsmath,graphicx,bm}
\usepackage{xcolor}
\usepackage{cleveref}

\newcommand\intraA{$d_{xy}$}
\newcommand\intraB{$d_{xz}~,~d_{yz}$}
\newcommand\interA{$d_{xz}/d_{yz}$}
\newcommand\interB{$d_{xy}/d_{xz}~,~d_{xy}/d_{yz}$}
\newcommand\sep{~:~}
\renewcommand\a{\alpha}
\renewcommand\b{\beta}
\newcommand\g{\gamma}
\newcommand\av{\mathbf{a}}
\newcommand\bv{\mathbf{b}}
\newcommand\cv{\mathbf{c}}
\newcommand\dv{\mathbf{d}}

\newcommand\dg{\dagger}

\newcommand\dn{\downarrow}
\newcommand\eps{\varepsilon}

\newcommand\Hc{\mathrm{H.c.}}

\newcommand\kv{\mathbf{k}}

\newcommand\R{\rangle}
\newcommand\rv{\mathbf{r}}
\newcommand\s{\sigma}

\newcommand\up{\uparrow}

\renewcommand\L{\langle}

\newcommand\SRO{Sr$_2$RuO$_4$}
\newcommand\ax{\mathbf{\hat a}_x}
\newcommand\ay{\mathbf{\hat a}_y}
\newcommand\az{\mathbf{\hat a}_z}
\newcommand\bx{\mathbf{\hat b}_x}
\newcommand\by{\mathbf{\hat b}_y}
\newcommand\bz{\mathbf{\hat b}_z}
\newcommand\cx{\mathbf{\hat c}_x}
\newcommand\cy{\mathbf{\hat c}_y}
\newcommand\cz{\mathbf{\hat c}_z}
\newcommand\dx{\mathbf{\hat d}_x}
\newcommand\dy{\mathbf{\hat d}_y}
\newcommand\dz{\mathbf{\hat d}_z}
\newcommand\ds{\mathbf{\hat d}_0}

\begin{document}
\title{Group-theoretic classification of superconducting states of Strontium Ruthenate}

\author{S.-O. Kaba}\affiliation{D\'epartement de physique and Institut quantique, Universit\'e de Sherbrooke, Sherbrooke, Qu\'ebec, Canada J1K 2R1}
\author{D. S\'en\'echal}\affiliation{D\'epartement de physique and Institut quantique, Universit\'e de Sherbrooke, Sherbrooke, Qu\'ebec, Canada J1K 2R1} 
\date{\today}

\begin{abstract}
The possible superconducting states of strontium ruthenate (\SRO) are organized into irreducible representations of the point group $D_{4h}$, with a special emphasis on nodes occurring within the superconducting gap. 
Our analysis covers the cases with and without spin-orbit coupling and takes into account the possibility of inter-orbital pairing within a three-band, tight-binding description of \SRO. 
No dynamical treatment if performed: we are confining ourselves to a group-theoretical analysis. 
The case of uniaxial deformations, under which the point group symmetry is reduced to $D_{2h}$, is also covered. It turns out that nodal lines, in particular equatorial nodal lines, occur in most representations.
We also highlight some results specific to multiorbital superconductivity. Among other things, we find that odd inter-orbital pairing allows to combine singlet and triplet superconductivity whithin the same irreducible representation, that pure inter-orbital superconductivity leads to nodal surfaces and that the notion of nodes imposed by symmetry is not clearly defined.
\end{abstract}
\maketitle

\section{Introduction}

The problem of identifying the symmetry of the superconducting order parameter in \SRO\ remains unsolved after more than 20 years \cite{review2003,mackenzie2017}. Despite the impressive number of experiments that were performed on high-quality samples, there is no clear consensus on the material's superconducting state. Initial NMR Knight shift \cite{knight}, neutron scattering \cite{neutron} and junction \cite{Nelson1151, laube2000, liu2010} experiments seemed to point towards triplet superconductivity, although this piece of evidence is now put to question by a recent study \cite{no-knight-shift}. There is also evidence for broken time reversal symmetry from muon spin resonance \cite{btrs} and polar Kerr effect \cite{kerr} measurements. These findings made plausible the early hypothesis of chiral triplet superconductivity \cite{order-parameter}, analogous to the A-phase of $\rm {}^3He$. However, some experiments are hard to conciliate with this scenario. First, specific heat and several transport probes showed the presence of residual excitations at low temperature \cite{specific-heat,taillefer,horizontal-nodes,lupien2001}, most likely related to gap nodes. Secondly, the presence of an effect resembling Pauli limiting must be present in the material to explain the value of $H_{c2}$ \cite{pauli-limiting}. Lastly, no splitting of the transition was observed when applying strain to the material\cite{hicks2014,pressure}.

Although strontium ruthenate shares a number of common characteristics with cuprates superconductors, among which its crystal structure \cite{maeno1994}, an important difference is its multiorbital nature. Its Fermi surface is well characterized and composed of three bands that have the character of Ru $t_{2g}$ orbitals. It is reasonable to believe that this fact plays an important, or at least a non-negligible, role in the superconductivity of this material. The identification of a dominant band for superconductivity in $\rm Sr_2RuO_4$ has not been unanimous \cite{orbital-dependant,mazin,hidden-1d,3b-rg-so}. Moreover, some studies suggest the possibility of important inter-orbital pairing in the material \cite{puetter2012,olivier}. This is not too surprising when considering that strong correlations arising in the material's normal state are mainly due to Hund's coupling \cite{yanase2014,acharya2017}, which is inter-orbital in nature. Spin-orbit coupling, which is also known to be significant in the material \cite{ng2000,apres2010,3b-quasiparticle,Tamai2019}, also has the effect to produce bands with mixed orbital character.

In the light of this situation, we propose to reexamine the different possibilities for the order parameter of~\SRO. A classification of possible order parameters must be done in terms of the irreducible representations of the point group symmetry of the lattice: $D_{4h}$, or $D_{2h}$ when uniaxial pressure is applied. This has already been done in previous works \cite{order-parameter,sigrist1996,sigrist1999}, but without fully considering the multiorbital nature of the material. This means that the order parameter must be considered not only as a space- and spin-dependent function, but also as an orbital-dependent function and that the irreducible representations are to be calculated accordingly. Such a classification is important, not only in order to frame all the proposals for superconducting order parameter in a coherent picture, but also because it can provide new insights about the superconducting state.
Note that we do not cover the possibility of odd-frequency pairing \cite{black-schaffer2013, geilhufe2018a, olivier} in the present work.

In this paper, we thus introduce a complete and rigorous classification of possible superconducting states in strontium ruthenate, akin to previous classifications that were made for high-temperature and heavy-fermions superconductors \cite{annett1990,sigrist1991}. We also highlight some features of multiorbital superconductivity that are different from what is seen in single-orbital superconductors. In particular, these considerations force us to rethink carefully about the relation between the spin character of the order parameter and its parity, the possibility of combining singlet and triplet superconductivity and the relation between order parameter symmetry and gap nodes. This classification also potentially applies to any $t_{2g}$ superconductor sharing the symmetry group of $\rm Sr_2RuO_4$.

This paper is organized as follows: In Sect.~\ref{sec:model} we introduce the tight-binding model used to describe \SRO\ and enumerate its symmetries.
In Sect.~\ref{sec:sym}, the main section of this paper, we explain how to classify the possible superconducting states into irreducible representations of the point group $D_{4h}$, with en emphasis on the existence or not of nodes in the gap. Possible pairing functions are listed in tables \ref{table:gapS}, \ref{table:gapT} and \ref{table:gapSO}, and generic nodes are illustrated on Fig.~\ref{fig:nodes}, and on Fig.~\ref{fig:nodes_D2h} in the case of uniaxial deformation. We offer some discussion and conclude in Sect.~\ref{sec:discussion}. This work is based on the Master's thesis of one of the authors~\cite{kaba2018}.

\section{The tight-binding model and its symmetries}\label{sec:model}

In this section we describe the Hamiltonian and its symmetries. We work in the orbital basis, not the band basis, even in reciprocal space, because it is the most appropriate to discuss symmetries.

\subsection{Hamiltonian}

\begin{figure}
\includegraphics[scale=0.8]{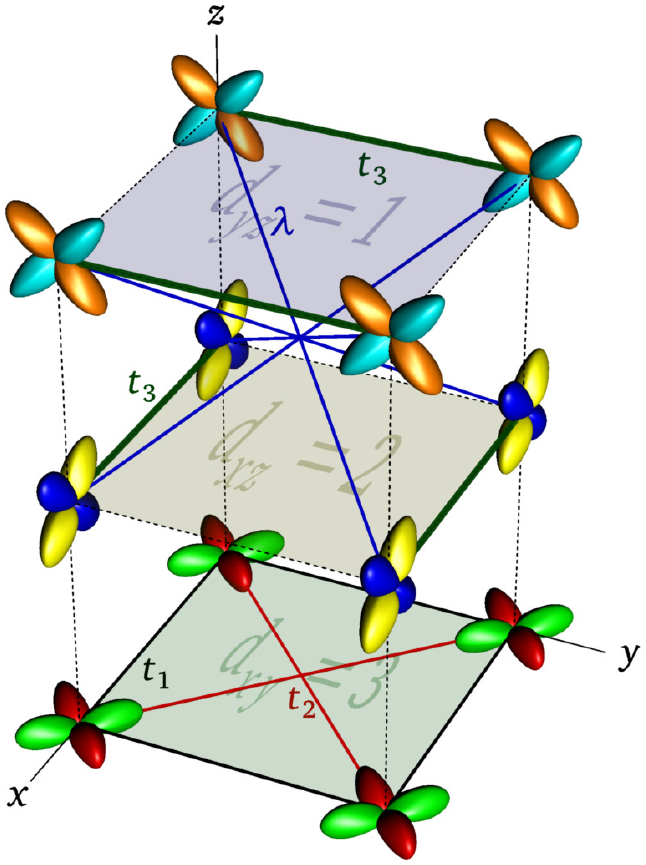}
\caption{
Schematic view of the SRO unit cell. The three orbitals have been vertically separated for clarity (the model considered is purely two-dimensional). the labels 1,2,3 correspond respectively to the $d_{yz}$, $d_{xz}$ and $d_{xy}$ orbitals. The different hopping terms ($t_{1,2,3}$ and $\lambda$) are illustrated.
}
\label{fig:SRO_cluster}
\end{figure}

We will assume that \SRO\ may be appropriately described by the following tight-binding, three-band Hamiltonian:
\begin{widetext}
\begin{multline}\label{eq:H0}
H_0 = t_1\sum_{\L \rv,\rv'\R,\s}  c^\dg_{\rv,3,\s}c_{\rv',3,\s} 
+ t_2 \sum_{\L \rv,\rv'\R_2,\s}  c^\dg_{\rv,3,\s}c_{\rv',3,\s} 
+ t_3 \left[\sum_{\L \rv,\rv'\R_x,\s} c^\dg_{\rv,1,\s}c_{\rv',1,\s}
 + \sum_{\L \rv,\rv'\R_y,\s}  c^\dg_{\rv,2,\s}c_{\rv',2,\s}\right]  \\
+ \lambda \sum_{\L \rv,\rv'\R_2,\s}  \left( c^\dg_{\rv,1,\s}c_{\rv',2,\s} + \Hc \right)
+ i\frac\kappa2 \sum_\rv\sum_{l,m,n}\eps_{lmn}
c_{\rv,l,\s}^\dg c_{\rv,m,\s'} \tau^n_{\s\s'} 
+ e\sum_{\rv, \s, m=1,2}  c^\dg_{\rv,m,\s}c_{\rv,m,\s} -\mu \sum_{\rv, m, \s}  c^\dg_{\rv,m,\s}c_{\rv,m,\s}
\end{multline}
where $c_{\rv,m,\s}$ is the annihilation operator for orbital $m=1,2,3$ of spin projection $\s$ at site $\rv$; $\L \rv,\rv'\R$ stands for nearest-neighbor pairs and $\L \rv,\rv'\R_2$ for second (diagonal) neighbors; $\L \rv,\rv'\R_x$ stands for nearest-neighbor
pairs in the $x$ direction, and likewise for the $y$ direction.
The $\kappa$ term is a spin-orbit coupling, where $\tau^{1,2,3}$ are the Pauli matrices and $\eps_{lmn}$ the Levi-Civita antisymmetric symbol.
Note that the chosen labeling of the three orbitals ($d_{yz}\to1$, $d_{xz}\to2$, $d_{xy}\to3$) is important in this expression.
Fig.~\ref{fig:SRO_cluster} illustrates the orbitals and hopping terms involved ($t_{1,2,3}$ and $\lambda$). 
On that figure, the three orbitals have been separated vertically for clarity.
The first two orbitals (1 and 2) are separated by an energy $e$ from the third.

The interaction terms include local Coulomb interactions $U$ (intra-orbital) and $U'$ (inter-orbital), as well as Hund couplings
$J$ and $J'$:
\begin{equation}
H_1 = \sum_{\rv} \left\{ U \sum_l n_{\rv,l,\up}n_{\rv,l,\dn} + 
\sum_{m\ne m'}\left[U'\sum_{\s,\s'} n_{\rv,m,\s} n_{\rv,m',\s'} 
+ \frac J2\sum_{\s,\s'} c^\dg_{\rv,m,\s}c^\dg_{\rv,m',\s'}c_{\rv,m,\s'}c_{\rv,m',\s}
+ \frac{J'}2\sum_{\s\ne\s'} c^\dg_{\rv,m,\s}c^\dg_{\rv,m,\s'}c_{\rv,m',\s'}c_{\rv,m',\s}
\right]\right\}
\end{equation}

The noninteracting Hamiltonian \eqref{eq:H0} can be expressed in momentum space:
\begin{equation}
H_0 = \sum_{m,m',\s,\s',\kv} c^\dg_{\kv,m,\s} \mathcal{H}_0(\kv)_{m\s, m'\s'} c_{\kv,m',\s'}
\end{equation}
with the $6\times6$ matrix
\begin{equation}
\mathcal{H}_0(\kv) = 
\begin{pmatrix}
e-\mu -2t_3\cos k_y & \frac i2\kappa + \lambda_\kv & 0 & 0 & 0 & -\frac12\kappa \\
-\frac i2\kappa + \lambda_\kv & e-\mu -2t_3\cos k_x & 0 & 0 & 0 & \frac i2\kappa \\
 0 & 0 & t_{1,\kv}+t_{2,\kv} -\mu & \frac12\kappa & -\frac i2\kappa & 0 \\
 0 & 0 & \frac12\kappa & e-\mu -2t_3\cos k_y & -\frac i2\kappa+\lambda_\kv & 0 \\
 0 & 0 & \frac i2\kappa & \frac i2\kappa+\lambda_\kv & 0 & 0 \\
 -\frac12\kappa & -\frac i2\kappa & 0 & 0 & 0 & t_{1,\kv}+t_{2,\kv} -\mu 
\end{pmatrix}
\end{equation}
\end{widetext}
where we have introduced
\begin{align}
t_{1,\kv} &= -2t_1(\cos k_x+\cos k_y) \\
t_{2,\kv} &= 4t_2\cos k_x\cos k_y \\
\lambda_\kv &= -4\lambda \sin k_x\sin k_y
\end{align}
The degrees of freedom are placed in the following order:
\begin{equation}
(1\up, 2\up, 3\up, 1\dn, 2\dn, 3\dn)
\end{equation}

Upon diagonalizing the matrix $\mathcal{H}_0(\kv)$ when $\kappa=0$, one recovers three bands:
The $d_{xy}$ orbital forms a band of its own labeled $\g$; the other two orbitals hybridize because of the $\lambda$ term and form two bands labeled $\a$ and $\b$. The associated Fermi surfaces are illustrated in red on Fig.~\ref{fig:nodes_descr}.
However, symmetries are more easily described in terms of the original orbitals, and therefore we will stick to the orbital description in the remainder of this paper.

Throughout this work, we use the following values of the band parameters: $t_1 = 1$ (the unit of energy), $t_2=0.4$, $t_3=1.25$, $\lambda = -0.1$, $e=0.1$ and $\mu=1.5$. 
These values are compatible with the ones used in the literature~\cite{ferro-vs-anti, photoemission, orbit+spin-fluct, 3b-rg-so, 3b-rg}.
When present, the spin-orbit coupling $\kappa$ is set to 0.2; this value was chosen somewhat arbitrarily, in order to have a visible impact on the dispersion relation (or Fermi surface).

\subsection{Symmetries}

The Hamiltonian $H=H_0+H_1$ has the following symmetries:
\begin{enumerate}
\item A mirror symmetry $\s_x$ with respect to the $yz$ plane; this reflexion changes the signs of orbitals 2 and 3.
\item A mirror symmetry $\s_y$ with respect to the $xz$ plane; this reflexion changes the signs of orbitals 1 and 3.
\item A $\pi/2$ rotation $C_4$ around the $z$ axis, together with the following exchange of orbitals:$d_{xz}\to d_{yz}$ and $d_{yz}\to -d_{xz}$ orbitals. The $d_{xy}$ orbital changes sign under this rotation.
\item Even though the model is two-dimensional, we could imagine a reflexion $\s_z$  with respect to the $xy$ plane that changes the signs of orbitals 2 and 3. Strictly speaking, this is an internal symmetry in the context of a two-dimensional model, but it will be relevant when classifying the superconducting pairing functions.\\ 
Operations 1-4 above generate the 16-element point group $D_{4h}$.
\item If $\kappa=0$, a simultaneous rotation of all spins. Otherwise these rotations are not independent of the spatial symmetries (see below).
\item Time-reversal
\item A $U(1)$ symmetry leading to the conservation of the total number of electrons in all three orbitals.
\item A $Z_2$ symmetry leading to the separate conservation of the parity (odd or even) of the number of electrons (i) in the $d_{xy}$ orbitals and (ii) in the $d_{yz}$ and $d_{xz}$ orbitals. Indeed, were it not for the interactions, the number of electrons would be separately conserved in the $d_{xy}$ on the one hand, and in the set $d_{yz}$,$d_{xz}$ on the other hand. The Hund coupling, however, allows pair hopping between these two sets.
\item Translation symmetry on the lattice.
\end{enumerate}

Let us consider a symmetry transformation $g$ acting on space.
In the absence of spin-orbit coupling, such a transformation does not affect spin and its effect on the annihilation operator $c_{\rv,m,\s}$ is the following:
\begin{equation}
	c_{\rv,m,\s} \to c'_{\rv,m,\s} = \sum_{m'} U_{mm'}(g) c_{g\rv,m',\s}
\end{equation}
where $g\rv$ is the mapping of site $\rv$ under the spatial symmetry transformation and $U(g)$ is a $3\times3$ matrix.
On the other hand, when $\kappa\ne0$, such a transformation must be accompanied by a spin rotation:
\begin{equation}
	c_{\rv,m,\s} \to c'_{\rv,m,\s} = \sum_{m',\s'} S_{\s\s'}(g) U_{mm'}(g) c_{\rv',m',\s'}
\end{equation}
Under this more general transformation, the spin-orbit term becomes
\begin{equation}
i\frac\kappa2 \sum_\rv\sum_{l,m,n}\eps_{l'm'n}U^*_{l'l}U_{m'm} 
c_{\rv,l,\s}^\dg c_{\rv,m,\s'} S^*_{\a\s}S_{\a'\s'}\tau^n_{\a\a'} 
\end{equation}
The spin rotation matrix $S$ must belong to a spinorial representation of the group such that
\begin{equation}\label{eq:spinrot}
\begin{aligned}
S^\dg \tau^n S &= R_{nn'}\tau^{n'} \\
\eps_{l'm'n}U^*_{l'l}U_{m'm} &= R^{-1}_{nn'}\eps_{lmn'}
\end{aligned}
\end{equation}
in order for the spin-orbit term to be invariant.
	
\begin{table}
\caption{
Generators of the point group $D_{4h}$. $U$ is the orbital part, $S$ the spin part (in the case of spin-orbit coupling) and $R$ the associated rotation of the $\dv$ vector.\label{table:generators}
}
\begin{ruledtabular}
\begin{tabular}{CCCC}
\mbox{generator} & U & S & R\\[6pt]
C_4 & \begin{pmatrix}0 & 1 & 0\\ -1 & 0 & 0 \\ 0 & 0 &-1\end{pmatrix} 
& \begin{pmatrix}\frac{1+i}{\sqrt2} & 0 \\ 0 & \frac{1-i}{\sqrt2}\end{pmatrix}
& \begin{pmatrix}0 & 1 & 0\\ -1 & 0 & 0 \\ 0 & 0 &1\end{pmatrix}	\\[14pt]
\s_x & \begin{pmatrix} 1 & 0 & 0\\ 0 & -1 & 0 \\ 0 & 0 & -1\end{pmatrix} 
& \begin{pmatrix}0 & i \\ i & 0\end{pmatrix}
& \begin{pmatrix} 1 & 0 & 0\\ 0 & -1 & 0 \\ 0 & 0 & -1\end{pmatrix}  \\[14pt]
\s_y & \begin{pmatrix} -1 & 0 & 0\\ 0 & 1 & 0 \\ 0 & 0 & -1\end{pmatrix} 
& \begin{pmatrix}0 & 1 \\ -1 & 0\end{pmatrix} 
& \begin{pmatrix} -1 & 0 & 0\\ 0 & 1 & 0 \\ 0 & 0 & -1\end{pmatrix} \\[14pt]
\s_z & \begin{pmatrix} -1 & 0 & 0\\ 0 & -1 & 0 \\ 0 & 0 & 1\end{pmatrix} 
& \begin{pmatrix}i & 0 \\ 0 & -i\end{pmatrix} 
& \begin{pmatrix} -1 & 0 & 0\\ 0 & -1 & 0 \\ 0 & 0 & 1\end{pmatrix} 
\end{tabular}
\end{ruledtabular}
\end{table}

The appropriate matrices $U$ and $S$, as well as the resulting rotation matrix $R$ of Eq.~\eqref{eq:spinrot}, are listed in Table~\ref{table:generators} for the four generators $C_4$, $\s_x$, $\s_y$ and $\s_z$. One checks that these matrices guarantee the invariance of the complete Hamiltonian under these transformations; in the absence of spin-orbit coupling, one can simply ignore the last two columns.
These three transformations generate a group isomorphic to $D_{4h}$, which as 16 elements and 10 irreducible representations (or \textit{irreps}, as we will call them from now on). 
Its character table is given on Table~\ref{table:D4h}.
The precise form of the $U$ matrices takes into account the change of sign of the $d$-orbitals under spatial transformations.

\begin{table*}
\caption{
Character table of $D_{4h}$, with a list of the simplest (i.e., lowest degree) singlet and triplet pairing functions for a single-band model without spin-orbit representation. Note that we distinguish $g$-type representations, which are even under inversion ($i$), from $u$-type representations, which are odd under inversion.\label{table:D4h}
}
\begin{ruledtabular}
\begin{tabular}{RRRRRRRRRRRRlr} 
& E & 2C_4 & C_2 & 2C_2' & 2C_2'' & i & 2S_4 & \sigma_z & \sigma_{x,y} & \sigma_{d,d'} & \mbox{pairing function} & nodes & spin
\\[4pt] \hline\\[-6pt]
A_{1g} & 1 & 1 & 1 & 1 & 1 & 1 & 1 & 1 & 1 & 1       &  1 & none  & 0 \\ 
A_{2g} & 1 & 1 & 1 & -1 & -1 & 1 & 1 & 1 & -1 & -1   &  x y (x^2 - y^2) & 8-fold  & 0 \\ 
B_{1g} & 1 & -1 & 1 & 1 & -1 & 1 & -1 & 1 & 1 & -1   &  x^2 - y^2 & 4-fold diagonal  & 0 \\ 
B_{2g} & 1 & -1 & 1 & -1 & 1 & 1 & -1 & 1 & -1 & 1   &  x y & 4-fold  & 0 \\ 
E_g & 2 & 0 & -2 & 0 & 0 & 2 & 0 & -2 & 0 & 0        &  z(x,y) & equator, 2-fold  & 0 \\ 
A_{1u} & 1 & 1 & 1 & 1 & 1 & -1 & -1 & -1 & -1 & -1  &  x y z (x^2 - y^2) & equator, 8-fold  & 1 \\ 
A_{2u} & 1 & 1 & 1 & -1 & -1 & -1 & -1 & -1 & 1 & 1  &  z & equator  & 1 \\ 
B_{1u} & 1 & -1 & 1 & 1 & -1 & -1 & 1 & -1 & -1 &  1 &  x y z & equator, 4-fold  & 1 \\ 
B_{2u} & 1 & -1 & 1 & -1 & 1 & -1 & 1 & -1 &  1 & -1 &  z (x^2 - y^2) & equator, 4-fold diagonal  & 1 \\ 
E_u & 2 & 0 & -2 & 0 & 0 & -2 & 0 & 2 & 0 & 0        &  (x,y) &  2-fold ($\mathbb{R}$) or none ($\mathbb{C}$)  & 1 
\end{tabular}
\end{ruledtabular}
\end{table*}

\section{Symmetries of the order parameter}\label{sec:sym}

\subsection{General considerations}

A general superconducting order parameter may be expressed in real space as
\begin{equation}
\Delta_{\rv,m,\s;\rv',m',\s'} = \L c_{\rv,m,\s} c_{\rv',m',\s'} \R
\end{equation}
Assuming translation symmetry, this order parameter depends on the difference $\rv-\rv'$ and is diagonal in $\kv$-space:
\begin{equation}
\Delta_{m,\s;m',\s'}(\kv) = \L c_{m,\s}(\kv) c_{m',\s'}(-\kv) \R~~.
\end{equation}
The Pauli principle imposes antisymmetry under the exchange of the quantum numbers of the pair:
\begin{equation}
	\Delta_{m,\s;m',\s'}(\kv) = -\Delta_{m',\s';m,\s}(-\kv)~~.
\end{equation}
In the remainder of this paper, the words \textit{symmetric} and \textit{antisymmetric} will refer to the properties of various parts of the pairing function with respect to the exchange of the two electrons.

A general order parameter function (or pairing function) can be expressed as a linear combination of basis functions. We can use a basis made of tensor products of position-dependent, orbital-dependent and spin-dependent factors:
\begin{equation}\label{eq:general_gap}
\Delta_{m,\s;m',\s'}(\kv) = \sum_{\mu\nu\rho} C_{\mu\nu\rho} f^\mu(\kv) O^\nu_{mm'} S^\rho_{\s\s'}~~.
\end{equation}


The spin part of the pairing function is generally described by the so-called $\dv$-vector, defined as follows:
\begin{equation}
S_{\s\s'} = i d_\rho (\tau_\rho\tau_2)_{\s\s'} = d_\rho \mathbf{\hat d}_\rho~~.
\end{equation}
The three components $d_{x,y,z}$ form the symmetric, triplet part of the spin part of the pairing function, whereas the antisymmetric, singlet part is represented by the zeroth component $d_0$ (the set of Pauli matrices $\tau_{1,2,3}$ is augmented by the identity matrix $\tau_0$).
Under a rotation in spin space, the three-vector $\dv$ transforms as a pseudo-vector (i.e., invariant under inversion), and $d_0$ behaves like a pseudo-scalar (it changes sign under inversion). In the presence of spin-orbit coupling, $d_z$ falls into the $A_{2g}$ representation and $(d_x,d_y)$ into $E_g$, whereas $d_0$ corresponds to $A_{1g}$.


Likewise, we will define the following $3\times3$ matrices to serve as a basis in orbital space:
\begin{align*}
\ax &= \begin{pmatrix} 1 & 0 & 0 \\ 0 & 0 & 0 \\ 0 & 0 & 0 \end{pmatrix} ~
&\bx &= \begin{pmatrix} 0 & 0 & 0 \\ 0 & 0 & 1 \\ 0 & 1 & 0 \end{pmatrix} ~
&\cx &= \begin{pmatrix} 0 & 0 & 0 \\ 0 & 0 & 1 \\ 0 & -1 & 0 \end{pmatrix}\\
\ay &= \begin{pmatrix} 0 & 0 & 0 \\ 0 & 1 & 0 \\ 0 & 0 & 0 \end{pmatrix} ~
&\by &= \begin{pmatrix} 0 & 0 & 1 \\ 0 & 0 & 0 \\ 1 & 0 & 0 \end{pmatrix} ~
&\cy &= \begin{pmatrix} 0 & 0 & 1 \\ 0 & 0 & 0 \\ -1 & 0 & 0 \end{pmatrix}\\
\az &= \begin{pmatrix} 0 & 0 & 0 \\ 0 & 0 & 0 \\ 0 & 0 & 1 \end{pmatrix} ~
&\bz &= \begin{pmatrix} 0 & 1 & 0 \\ 1 & 0 & 0 \\ 0 & 0 & 0 \end{pmatrix} ~
&\cz &= \begin{pmatrix} 0 & 1 & 0 \\ -1 & 0 & 0 \\ 0 & 0 & 0 \end{pmatrix}
\end{align*}
A general matrix acting on orbital space may then be expressed via three
vectors $\av$, $\bv$ and $\cv$ as
\begin{equation}
	O_{mn} = \av\cdot\mathbf{\hat a}_{mn} + \bv\cdot\mathbf{\hat b}_{mn} + \cv\cdot\mathbf{\hat c}_{mn}~~.
\end{equation}
The components of these vectors, like the annihilation operators $c_{\rv,m,\s}$, will be labeled using indices $m=x,y,z$, corresponding respectively to the three orbitals $d_{yz}$, $d_{xz}$ and $d_{xy}$, also numbered $1,2,3$ in Fig.~\ref{fig:SRO_cluster}.
Clearly the $\av$ and $\bv$ vectors describe symmetric orbital parts of the pairing function, and $\cv$ antisymmetric orbital parts.
The advantage of defining the vectors $\av$, $\bv$ and $\cv$ lies in their transformation properties: 
The combinations $a_z$ and $a_x+a_y$ belong to the $A_{1g}$ representation, and $a_x-a_y$ to $B_{1g}$.
The component $b_z$ belongs to $B_{2g}$, whereas $c_z$ belongs to $A_{2g}$. Finally, the pairs $(b_x,b_y)$ and $(c_x,c_y)$ both belong to $E_g$.
Said differently, the $\av$ vector transforms like the functions $(x^2,y^2,z^2)$, the $\bv$ vector like the functions $(yz,xz,xy)$ and the $\cv$ vector like a pseudo-vector.


As for the spatial part of the pairing function, it will be described by multinomials in $x,y,z$, which in fact stand for the components $k_x,k_y,k_z$ of the wavevector.
The three linear functions $\{x,y,z\}$ form a ``vector'' representation of $D_{4h}$, which is obviously reducible: $z$ belongs to the $A_{2u}$ representation, and $(x,y)$ form the two-component $E_u$ representation. By taking symmetrized tensor products of this reducible representation repeatedly with itself, one finds reducible representations for quadratic, cubic, quartic functions, and so on.
The even-degree functions are symmetric under inversion (which corresponds here to exchanging the spatial quantum numbers), whereas the odd-degree functions are antisymmetric.

The coefficient $C_{\mu\nu\rho}$ of Eq.~\eqref{eq:general_gap} will therefore be expressed in terms of components of the $\dv$-vector for the spin index $\rho$, components of the $\av$, $\bv$ and $\cv$ vectors for the orbital index $\nu$, and multinomial functions of $x,y,z$ for the spatial part. For instance, the pairing function $\cz  \dz (x^2-y^2)$, which appears below in Table~\ref{table:gapSO} under the $B_{1g}$ representation, represents a spin triplet $(|\!\!\up\dn\rangle+|\!\!\dn\up\rangle)$ with an antisymmetric orbital combination of $d_{xz}$ and $d_{yz}$ (because of $\cz$), and a $d$-wave-like spatial part. The product $\cz\dz$ is a tensor product of a $3\times3$ matrix acting in orbital space ($\cz$) with a $2\times2$ matrix acting in spin space ($\dz$), so that the overall pairing function in this case is a $6\times6$ matrix.

	\subsubsection{Landau theory}

	We assume that the pattern of symmetry breaking occurs within the framework of the Landau theory of phase transitions. A generic superconducting order parameter may be decomposed on a basis of possible pairing functions $\hat\Delta_\mu$, i.e., $\Delta = \sum_\mu \psi_\mu \hat\Delta_\mu$, and the Landau free energy functional is a power expansion in terms of the coefficients $\psi_\mu$:
	\begin{equation}
	f[\psi] = a_{\mu\nu}(T)\psi_\mu^*\psi_\nu + b_{\mu\nu\rho\lambda}(T)\psi_\mu^*\psi_\nu^*\psi_\rho\psi_\lambda + \cdots
	\end{equation}
	where the ellipsis stands for gradient and higher-degree terms, and $T$ is the temperature.
	
	Organizing the basis functions $\hat\Delta_\mu$ according to irreps of the point group makes the matrix $a(T)$ block-diagonal: $a(T) = \bigoplus_r a^{(r)}(T)$, i.e., it has no matrix elements between functions belonging to different irreps.
	Within each representation, the matrix $a^{(r)}(T)$ may be diagonalized, and at some point upon lowering $T$ one of its eigenvalues, initially all positive, may change sign, which signals the superconducting phase transition and a minimum of $f[\psi]$ at $\psi\ne0$.
	This is going to first occur in one of the representations and will define the symmetry character of the superconducting state. Nothing forbids competing minima, and hence additional phase transitions, to appear at lower temperatures. These transitions should be detectable, for instance by specific heat measurements. None has been seen in \SRO~\cite{review2003, mackenzie2017}, and therefore we will assume a single symmetry breaking pattern in this work.
	
	If the transition occurs in the $A_{1g}$ representation, then the only broken symmetry is the $U(1)$ of gauge invariance. In any other irrep, the point group $D_{4h}$ is broken as well, but not completely: The minimum $\psi^\star$ leaves a subgroup of $D_{4h}$ invariant.
	For instance, in the $B_{1g}$ representation, the superconducting state is effectively a distortion that breaks $D_{4h}$ down to the group $D_{2h}$ as described in Sect.~\ref{sec:D2h} below, and it happens that all basis functions of $B_{1g}$ are invariant under this subgroup.
	It is noteworthy that for a group like $D_{4h}$, which only has one-dimensional and two-dimensional chiral-like irreps ($E_g$ and $E_u$), this invariant subgroup only depends on the irrep of the solution, i.e., it is the same for all basis functions within that irrep.
	This means that, in a given state of broken symmetry, all basis functions of a given irrep may a priori contribute to the total (or combined) pairing function.
	
	Time reversal (TR) symmetry may only occur when the minimum $\psi^\star$ is degenerate, and this will occur only within a two-dimensional representation ($E_g$ or $E_u$). In those cases, the complex combination $(1,i)$ of the two basis functions defines a broken TR state, with the conjugate combination $(1,-i)$ being the time-reversed state. Other TR broken states could only occur when two solutions belonging to different representations happen to have the same energy, which implies a second phase transition as mentioned above. We exclude that possibility.
	
	\subsection{Quasi-Particle Dispersion}
	
	In order to identify nodes, or other elementary properties of the superconducting state, one must compute the quasi-particle dispersion; this is done at the mean-field level. 
	
	The pairing function $\Delta_\kv$ is a $6\times6$ matrix. 
	It appears in the mean-field Hamiltonian as
	\begin{equation}
	F = \sum_{m,m',\s,\s',\kv} c_{\kv,m,\s} \Delta(\kv)_{m\s, m'\s'} c_{-\kv,m',\s'} + \mbox{H.c.}~~.
	\end{equation}
	The normal and anomalous part of the Hamiltonian are put together via
	the Nambu formalism, in which we introduce a 12-component spinor at a given wavevector $\kv$:\footnote{Note here that we did not invert the spin quantum number in the second half of the Nambu spinor. This is matter of convenience and amounts to changing the order of the components compared to the usual convention.}
	\begin{equation}
	\Psi_\kv = \left( c_{\kv,m,\s} , c^\dg_{-\kv,m,\s} \right)
	\qquad (m=1,2,3;~\s=\up,\dn)~~.
	\end{equation}
	The combined Hamiltonian takes the following form:
	\begin{equation}
	H = \sum_\kv \Psi^\dg_\kv \mathcal{H}(\kv) \Psi_\kv
	\end{equation}
	with the $12\times12$ matrix
	\begin{equation}
	\mathcal{H}(\kv) = \begin{pmatrix}
	\mathcal{H}_0(\kv) & \Delta(\kv) \\
	\Delta^\dg(\kv) & -\mathcal{H}_0^*(\kv)
	\end{pmatrix}~~.
	\end{equation}
	The eigenvalues of $\mathcal{H}(\kv)$ occur in pairs of opposite signs and provide the dispersion relation of the quasiparticles. Nodes are found by looking for the zeros of these eigenvalues.
		
\begin{table*}
\caption{
List of singlet pairing functions (no spin-orbit coupling). Functions are arranged
according to $D_{4h}$ representations and type of inter-orbital pairing: intra-orbital ($d_{xy}$, $d_{xz}$ or $d_{yz}$) or inter-orbital ($d_{xy}/d_{xz}$, $d_{xy}/d_{yz}$, $d_{xy}/d_{yz}$, $d_{xz}/d_{yz}$).
The notation for the nodes is the following: $\a$, $\b$ and $\g$ refer to the normal state Fermi surface sheets (see Fig.~\ref{fig:nodes_descr}); when appearing alone, it means that the whole sheet is a nodal surface. They can be hybridized, hence the notation $(\b\g)$, etc. When appearing next to a symbol ($+$, $|$ , $-$, $\times$, $\diagdown$), then the node is the intersection of that sheet with particular lines: $+$ stands for horizontal and vertical axes at 0 and $\pm\pi$; $-$ and $|$ stand for horizontal and vertical lines only, whereas $\times$ stands for the diagonals and $\diagdown$ for the north-west diagonal only. Commas separate different nodal lines or surfaces present. 
The combined nodes are obtained in mixing the different functions in a given representation.
For functions that do not involve $z$, the nodes are $k_z$ independent in our approximation that neglects hopping in the $z$ direction. For functions that depend on $z$, the nodes indicated here are computed at $k_z=\pi/2$; in those cases the pairing function vanishes at $k_z=0$ and the nodes there coincide with the complete Fermi surface.\label{table:gapS}
}

\begin{ruledtabular}
\begin{tabular}{lllLL}
irrep & combined nodes & orbital mixing & \mbox{pairing function} & \mbox{nodes}  \\ 
\hline\multirow{4}{*}{$A_{1g}$}
&  \multirow{4}{*}{none}
&  \intraA    & \az  & \a,\b \\ 
&&  \intraB  & \ax  + \ay  & \g \\
&&  \interA  & \bz  x y & \g, +\a, +\b \\ 
&&  \interB  & z (\bx  y - \by  x) & (\a\b\g) \\
\hline\multirow{4}{*}{$A_{2g}$} 
& \multirow{4}{*}{$+\g, +\b$}
& \intraA  & \az  x y (x^2 - y^2) & \a,\b,+\times\g \\
&& \intraB  & x y (\ax  - \ay ) & \g, +\times\a, +\b \\
&& \interA  & \bz  (x^2 - y^2) & \g, +\times\a, +\times\b \\ 
&& \interB  & z (\bx  x + \by  y) & \a,\b,\g \\
\hline\multirow{4}{*}{$B_{1g}$} 
& \multirow{4}{*}{$\times\b$}
& \intraA  & \az  (x^2 - y^2) & \times\g,\a,\b \\
&& \intraB  & \ax  - \ay & \g, \times\a, \times\b \\
&& \interA  & \bz  x y (x^2 - y^2) & \g, +\times\a, +\times\b \\ 
&& \interB  & z (\bx  y + \by  x) & (\a\g), \b \\
\hline\multirow{4}{*}{$B_{2g}$} 
& \multirow{4}{*}{$+\a,+\g,+\b$}
& \intraA  & \az  x y & \a,\b, +\g \\
&& \intraB  & x y (\ax  + \ay )& \g, +\a, +\b\\
&& \interA  & \bz & \g, +\a, +\b\\
&& \interB  & z (\bx  x - \by  y) & \a, (\b\g) \\
\hline\multirow{4}{*}{$E_{g}$} 
& \multirow{4}{*}{$\b,(\a\g)\sep(\a\b\g)\sep\b,(\a\g) $}
& \intraA  & \az  z (x,y) & \a,\b,|\g \sep \a,\b,\diagdown\g \sep \a,\b \\
&& \intraB  & z(\ax  x, \ay  y) & +\a,+\b,\g\sep+\times\a,+\times\b,\g\sep +\a,+\b,\g \\
&&   & z(\ax  y, \ay  x) & -\a,-\b,\g\sep\diagdown\a\diagdown\b,\g\sep \g \\
&& \interA  & \bz  z(x,y) & +\a,+\b,\g\sep+\a,+\b,\g\sep +\a,+\b,\g \\
&& \interB  & (\bx ,\by ) & \b,(\a\g)\sep(\a\b\g)\sep \b,(\a\g) \\
\hline\multirow{2}{*}{$A_{1u}$} 
& \multirow{2}{*}{$\b,(\a\g)$}
& \interA  & \cz  z & \a,\b,\g \\ 
&& \interB  & \cx  x + \cy  y & \b,(\a\g) \\ 
\hline\multirow{2}{*}{$A_{2u}$} 
& \multirow{2}{*}{$\a,(\b\g)$}
& \interA  & \cz  x y z (x^2 - y^2) & \a,\b,\g \\ 
&& \interB  & \cx  y - \cy  x & (\a\b\g) \\
\hline\multirow{2}{*}{$B_{1u}$} 
& \multirow{2}{*}{$\a,(\b\g)$}
& \interA  & \cz  z (x^2 - y^2) & \a,\b,\g \\ 
&& \interB  & \cx  x - \cy  y & \a,(\b\g) \\
\hline\multirow{2}{*}{$B_{2u}$} 
& \multirow{2}{*}{$\b,(\a\g)$}
& \interA  & \cz  xyz & \a,\b,\g \\ 
&& \interB  & \cx  y + \cy  x & \b,(\a\g) \\
\hline\multirow{2}{*}{$E_{u}$} 
& \multirow{2}{*}{$\a,\b,\g\sep (\a\b\g)\sep\a,\b,\g$}
& \interA  & \cz  (x,y) & \a,\b,\g \sep \g,(\a\b) \sep (\a\b),\g\\ 
&& \interB  & z(\cx ,\cy ) & (\a\g),\b \sep (\a\b\g) \sep (\a\g),\b
\end{tabular}
\end{ruledtabular}
\end{table*}

\begin{table*}
\caption{
List of triplet pairing functions (no spin-orbit coupling). See Table.~\ref{table:gapS} and text for an explanation.\label{table:gapT}
}
\begin{ruledtabular}
\begin{tabular}{lllLL}
irrep & combined nodes & orbital mixing & \mbox{pairing function} & \mbox{nodes}  \\ 
\hline\multirow{2}{*}{$A_{1g}$} & \multirow{2}{*}{$\a,\b,\g$}
&  \interA  & \cz  x y (x^2 - y^2) & \a,\b,\g \\ 
&&  \interB  & z (\cx  y - \cy  x) & (\a\b\g) \\
\hline\multirow{2}{*}{$A_{2g}$} & \multirow{2}{*}{$\a,\b,\g$}
&  \interA  & \cz  & \a,\b,\g \\ 
&&  \interB  & z (\cx  x + \cy  y) & (\a\g),\b \\
\hline\multirow{2}{*}{$B_{1g}$} & \multirow{2}{*}{none}
&  \interA  & \cz  x y & \a,\b,\g \\ 
&&  \interB  & z (\cx  y + \cy  x) & (\a\g),\b \\
\hline\multirow{2}{*}{$B_{2g}$} & \multirow{2}{*}{none}
&  \interA  &  \cz  (x^2 - y^2) & \a,\b,\g \\ 
&&  \interB  & z (\cx  x - \cy  y) & \a,(\b\g) \\
\hline\multirow{2}{*}{$E_{g}$} & \multirow{2}{*}{$\b,(\a\g)\sep(\a\b\g)\sep\b,(\a\g)$}
&  \interA  & \cz  z(x,y) & \a,\b,\g\sep (\a\b),\g\sep (\a\b),\g \\ 
&&  \interB  & (\cx , \cy ) & \b,(\a\g)\sep (\a\b\g) \sep \b,(\a\g)\\
\hline\multirow{4}{*}{$A_{1u}$} & \multirow{4}{*}{$\b,(\a\g)$}
& \intraA  & \az  x y z (x^2 - y^2) & \a,\b,+\g \\
&& \intraB  & x y z (\ax  - \ay ) & +\a,+\times\b,\g \\
&& \interA  &  \bz  z (x^2 - y^2) & +\times\a,+\times\b,\g \\
&& \interB  &  \bx  x +  \by  y & \b,(\a\g) \\
\hline\multirow{4}{*}{$A_{2u}$} & \multirow{4}{*}{$\a,(\b\g)$}
& \intraA  & \az  z & \a,\b \\
&& \intraB  & z (\ax  + \ay ) & \g \\
&& \interA  & \bz  x y z & +\a,+\b,\g \\
&& \interB  & \bx  y -  \by  x & (\a\b\g) \\
\hline\multirow{4}{*}{$B_{1u}$} & \multirow{4}{*}{$\a,(\b\g)$}
& \intraA  & \az  x y z & \a,\b,+\g \\
&& \intraB  & x y z (\ax  + \ay ) & +\a,+\b,\g \\
&& \interA  & \bz  z & +\a,+\b,\g \\
&& \interB  & \bx  x -  \by  y & \a,(\b\g) \\
\hline\multirow{4}{*}{$B_{2u}$} & \multirow{4}{*}{$\b,(\a\g)$}
& \intraA  & \az  z (x^2 - y^2) & \a,\b,\times\g \\
&& \intraB  & z (\ax  - \ay ) & \times\a,\times\b,\g \\
&& \interA  &  \bz  x y z (x^2 - y^2) & +\times\a,+\times\b,\g \\
&& \interB  &  \bx  y +  \by  x & \b,(\a\g) \\
\hline\multirow{5}{*}{$E_{u}$} & \multirow{4}{*}{$|\g\sep\diagdown\g\sep$none} 
& \intraA  & \az (x,y) & \a,\b,|\g \sep  \a,\b,\diagdown\g \sep \a,\b \\
&& \intraB  & (\ax  x,\ay  y) & +\a,+\b,\g\sep +\a,+\b,\g \sep +\a,+\b,\g \\
&& & (\ax  y,\ay  x) & -\a,-\b,\g\sep \diagdown\a,\diagdown\b,\g \sep \g \\
&& \interA  & \bz (x,y) & +\a, +\b, \g \sep +\a,+\b,\g \sep +\a,+\b,\g \\
&& \interB  & z(\bx ,\by ) & \b,(\a\g) \sep (\a\b\g) \sep (\a\g),\b \\
\end{tabular}
\end{ruledtabular}
\end{table*}

\begin{table*}[p]
\caption{
List of pairing functions with spin-orbit coupling. See Table.~\ref{table:gapS} and text for an explanation.
\label{table:gapSO}}
\begin{ruledtabular}
\begin{tabular}{lllLL}
irrep & \mbox{combined nodes} & orbital mixing & \mbox{pairing function} & \mbox{nodes}  \\ 
\hline\multirow{4}{*}{$A_{1g}$} & \multirow{4}{*}{none} 
& \intraA   & \az  \ds  & \a,\b \\ 
&& \intraB   & \ds  (\ax  + \ay ) & \g \\ 
&& \interA   & \cz  \dz  & \a,\b,\g \\ 
&& \interB   & (\cx  \dx  - \cy  \dy ) & \a,\b,\g \\ 
\hline\multirow{4}{*}{$A_{2g}$} & \multirow{4}{*}{$+\a,+\b,+\g$} 
& \intraA   & \az  \ds  x y (-x^2 + y^2) & \a,\b,\times+\g \\ 
&& \intraB   & \ds  x y (-\ax  + \ay ) &  +\a,+\b,\g \\ 
&& \interA   & \cz  z (\dx  y + \dy  x)~,~ \bz  \ds  (x^2 - y^2) & \a,\b,\g \\ 
&& \interB   & \cx  \dy  + \cy  \dx  & \a,\b,\g \\ 
\hline\multirow{4}{*}{$B_{1g}$}& \multirow{4}{*}{$\times\g$} 
& \intraA   & \az  \ds  (x^2 - y^2) & \a,\b,\times\g \\ 
&& \intraB   & \ds  (\ax  - \ay ) & \g \\ 
&& \interA   & \cz  z (\dx  x + \dy  y) ~,~ \cz  \dz  (x^2 - y^2) & \a,\b,\g \\ 
&& \interB   & (\cx  \dx  + \cy  \dy ) & \a,\b,\g \\ 
\hline\multirow{4}{*}{$B_{2g}$} & \multirow{4}{*}{$+\a,+\b,+\g$} 
& \intraA   & \az  \ds  x y & \a,\b,+\g \\ 
&& \intraB   & \ds  x y (\ax  + \ay ) & +\a,+\b,\g \\ 
&& \interA   & \bz  \ds  & \a,\b,\g \\ 
&& \interB   & \cx  \dy  - \cy  \dx  & \a,\b,\g \\ 
\hline\multirow{6}{*}{$E_{g}$} &  
& \intraA   & \az  \ds  z(x,y) & \a,\b,|\g\sep \a,\b,\diagdown\g \sep \a,\b\\ 
&& \intraB   & \ds z(\ax  x ,\ay  y) & -\a,-\b,\g\sep +\a,+\b,\g \sep +\times\a,+\b,\g \\ 
& $|\b, |\g \sep$ &   & \ds z(\ax  y,\ay  x) & -\a,-\b,\g\sep \g \sep \times\a,\g \\ 
&$\diagdown\a,\diagdown\b,\diagdown\g$& \interA   & \cz (\dx , \dy ) & \a,\b,\g\sep\a,\b,\g \sep \a,\b,\g \sep\\ 
&$\diagdown\a,\diagdown\b$& \interB   & (- \by  \ds  +  \cy  \dz , \bx  \ds  +  \cx  \dz ) & \a,\b,\g\sep\a,\b,\g \sep \a,\b,\g \\ 
&& & (\by  \ds  +  \cy  \dz ,- \bx  \ds  +  \cx  \dz ) & \a,\b,\g\sep\a,\b,\g \sep \a,\b,\g \\ 
\hline\multirow{4}{*}{$A_{1u}$} & \multirow{4}{*}{$+\a,+\b,+\g$} 
& \intraA   & \az  \ds  x y z (x^2 - y^2) & \a,\b,+\times\g \\ 
&& \intraB   & \ds  x y z (\ax  - \ay ) & +\a,+\b,\g \\ 
&& \interA   & \cz  (\dx  y + \dy  x) & \a,\b,\g \\ 
&& \interB   & z (\cx  \dy  + \cy  \dx )~,~ \dz  (-\cx  y + \cy  x)~,~\ds  (\bx  x + \by  y) & \a,\b,\g \\ 
\hline\multirow{4}{*}{$A_{2u}$} & \multirow{4}{*}{none} 
& \intraA   & \az  \ds  z & \a,\b \\ 
&& \intraB   & \ds  z (\ax  + \ay ) & \g \\ 
&& \interA   & \cz  (\dx  x - \dy  y)~,~ \cz  \dz  z & \a,\b,\g \\ 
&& \interB   & z (\cx  \dx  - \cy  \dy )~,~\ds  (\bx  y - \by  x)~,~\dz  (\cx  x + \cy  y) & \a,\b,\g \\
\hline\multirow{4}{*}{$B_{1u}$} & \multirow{4}{*}{$+\a,+\b,+\g$} 
& \intraA   & \az  \ds  x y z & \a,\b,+\g \\ 
&& \intraB   & \ds  x y z (\ax  + \ay ) & +\a,+\b,\g \\ 
&& \interA   & \cz  (-\dx  y + \dy  x)~,~\bz  \ds  z & \a,\b,\g \\ 
&& \interB   & z (\cx  \dy  - \cy  \dx )~,~\dz  (\cx  y + \cy  x)~,~\ds  (\bx  x - \by  y) & \a,\b,\g \\ 
\hline\multirow{4}{*}{$B_{2u}$} & \multirow{4}{*}{none} 
& \intraA   & \az  \ds  z (x^2 - y^2) & \a,\b,\times\g \\ 
&& \intraB   & \ds  z (\ax  - \ay ) & \g \\ 
&& \interA   & \cz  (\dx  x + \dy  y) & \a,\b,\g \\ 
&& \interB   & z (\cx  \dx  + \cy  \dy )~,~ \ds  (\bx  y + \by  x)~,~\dz  (\cx  x - \cy  y) & \a,\b,\g \\ 
\hline\multirow{10}{*}{$E_{u}$} &
& \intraA   & \az  \ds  (x,y) & \a,\b,|\g \sep \a,\b,\diagdown\g \sep \a,\b \\ 
&& \intraB   & \ds  (\ax  x,\ay  y) & +\a,+\b,\g\sep +\a,+\b,\g\sep +\a,+\b,\g\\ 
&&& \ds  (\ax  y,\ay  x) & -\a,-\b,\g \sep \g\sep \times\a,\g\\ 
&& \interA   & \cz  z (\dx , \dy )~,~(\bz  \ds  - \cz  \dz )(x,y)~,~(\bz  \ds  + \cz  \dz )(x, y) &  \a,\b,\g \sep \a,\b,\g \sep \a,\b,\g \\ 
&$|\b, |\g\sep$& \interB   & ( \cy  x (\dx  - i \dy ),\cx  y (\dx  + i \dy )) & \a,\b,\g \sep\a,\b,\g \sep \a,\b,\g  \\ 
&$\diagdown\a,\diagdown\b,\diagdown\g\sep$&& ( \cy  y (\dx  - i \dy ),\cx  x (\dx  + i \dy )) & \a,\b,\g \sep\a,\b,\g \sep \a,\b,\g  \\ 
& none&& z(\by  \ds  - \cy  \dz , \bx  \ds  + \cx  \dz ) & \a,\b,\g \sep\a,\b,\g \sep \a,\b,\g \\ 
&&& z(\by  \ds  + \cy  \dz , \bx  \ds  - \cx  \dz ) & \a,\b,\g \sep\a,\b,\g \sep \a,\b,\g \\ 
&&& (\cy  x (\dx  + i \dy ), \cx  y (-\dx  + i \dy )) & \a,\b,\g \sep\a,\b,\g \sep\a,\b,\g  \\ 
&&& (\cy  y (\dx  + i \dy ), \cx  x (-\dx  + i \dy )) & \a,\b,\g \sep\a,\b,\g \sep\a,\b,\g  \\ 
\end{tabular}
\end{ruledtabular}
\end{table*}

\begin{figure}
\includegraphics[scale=0.9]{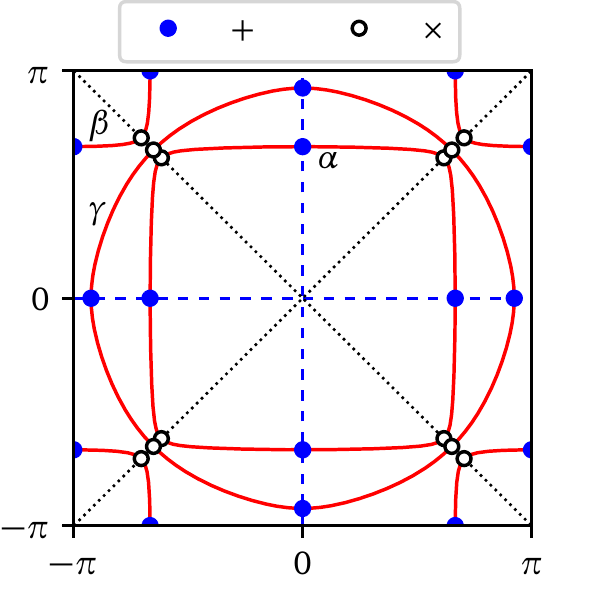}~~~
\caption{Illustration of the possible nodes in pairing functions.
In red: the normal state Fermi surface, with the band labels $\a$, $\b$ and $\g$.
In tables \ref{table:gapS}, \ref{table:gapT} and \ref{table:gapSO}, the $+$ sign stands for the intersection of any of these surfaces with horizontal/vertical axes (blue dots), including zone boundaries; The $\times$ sign stands for the intersection of either of these surfaces with diagonal axes (open black dots). The $-$, $|$ and $\diagdown$ signs will stand for the intersection with the horizontal, vertical and north-west diagonals only.
	}
	\label{fig:nodes_descr}
	\end{figure}

\subsection{No spin orbit coupling}

In the following, we will construct possible pairing functions $\Delta(\kv)$ organized according to irreps of the point group $D_{4h}$, keeping the spatial part as simple as possible.
The construction of pairing functions is simpler in the absence of spin-orbit coupling, because the spin part always factorizes from the rest and is either a singlet or a triplet. One can then concentrate on the construction of the spatial-orbital part, which must be symmetric in the singlet case, and antisymmetric otherwise. 

This construction can be automated as follows: One constructs a $3\times3$ matrix representation $U_{mm'}(g)$ of each of the 16 elements $g$ of $D_{4h}$ acting in orbital space, by combining the generators of Table~\ref{table:generators}. The symmetrized and antisymmetrized tensor products of this representation with itself are then constructed:
\begin{equation}
\mathcal{S}U(g)\otimes U(g) \quad\mbox{and}\quad \mathcal{A}U(g)\otimes U(g)
\end{equation}
($\mathcal{S}$ and $\mathcal{A}$ are the symmetrizer and antisymmetrizer, respectively).
The tensor products of these orbital representations with the spatial representations of a given degree in $(x,y,z)$ are constructed next. The resulting higher-dimensional representation $R(g)$ can then be projected onto irreps or $D_{4h}$ with the help of projection operators:
\begin{equation}
P^{(r)} = \frac{d^{(r)}}{|G|}\sum_{g\in G} \chi_g^{(r)*} R(g)
\end{equation}
where $G$ stands for the point group (here $D_{4h}$), the sum is over the $|G|$ group elements $g$, and $\chi^{(r)}$ is the character of the irrep $r$ (according to table \ref{table:D4h}).
This procedure is done using a combination of numerical and symbolic computations in the Python language.

Among the states selected by the projection operator, some involve only the vector $\av$ and therefore describe intra-orbital pairing. Those involving the components of $\bv$ describe inter-orbital pairing that is symmetric in orbital (and consequently associated to a symmetric spatial part for singlets and antisymmetric spatial part for triplets). Those involving the components of $\cv$ describe inter-orbital pairing that is antisymmetric in orbital (and consequently associated to an antisymmetric spatial part for singlets and symmetric spatial part for triplets)

Table \ref{table:gapS} lists the singlet pairing functions found in this way. They are enumerated according to irrep and, within each irrep, according to the type of orbital pairing: 
\begin{enumerate}
\item \intraA  : intra-orbital pairing within the $d_{xy}$ orbital, forming the so-called $\g$ band.
\item \intraB  : intra-orbital pairing within the $d_{yz}$ or $d_{xz}$ orbital.
\item \interA  : inter-orbital pairing between the $d_{yz}$ and $d_{xz}$ orbitals.
\item \interB  : inter-orbital pairing between $d_{xy}$  and $d_{yz}$ orbitals, or between $d_{xy}$ and $d_{xz}$ orbitals.
\end{enumerate}
For the sake of illustrating each type of orbital pairing, we have carried the construction of spatial functions to a degree sufficient to display all cases, but displaying only the lowest degree in each.
Column 4 of Table~\ref{table:gapS} shows the pairing function as a function of orbital vector and coordinates ($x,y,z$), or equivalently $(k_x,k_y,k_z)$. In order to represent lattice quantities in the full Brillouin zone and to identify nodes in the dispersion, we perform the following substitutions for $x$:
\begin{equation}
x\to\sin k_x \qquad\qquad x^2\to 1-\cos k_x 
\end{equation}
and likewise for $y$ and $z$.
Such a substitution would allows us to provide a real-space description of pairing. 
For instance, a product like $\sin k_x\sin k_y = \frac12[\cos(k_x-k_y) - \cos(k_x+k_y)]$ would correspond a cross-shaped pairing accross the nearest-neighbor diagonals, and so on.

\begin{figure*}[p]
	\includegraphics[scale=0.6]{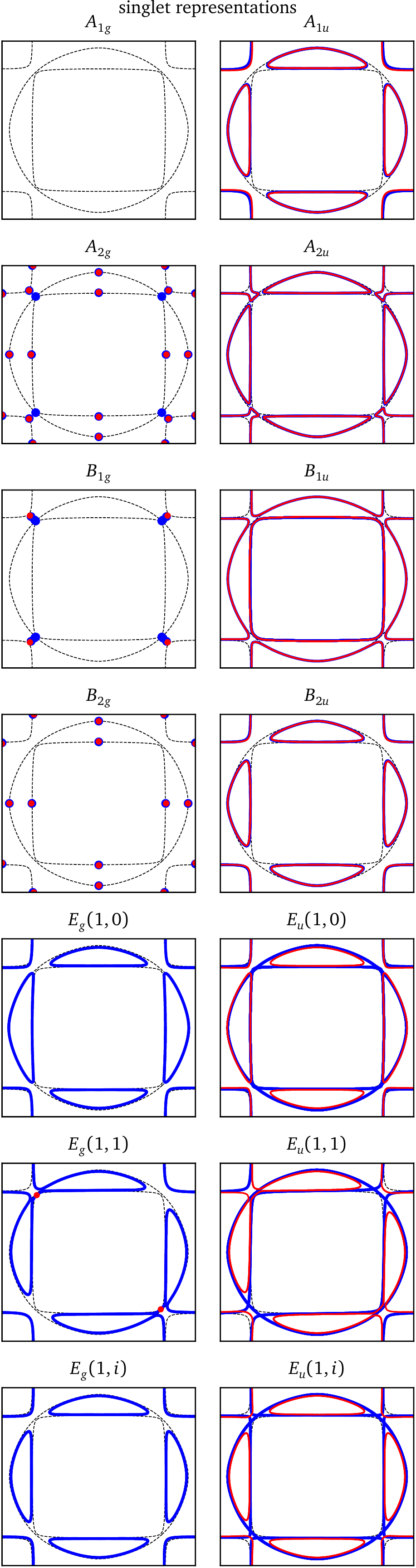}~~~
	\includegraphics[scale=0.6]{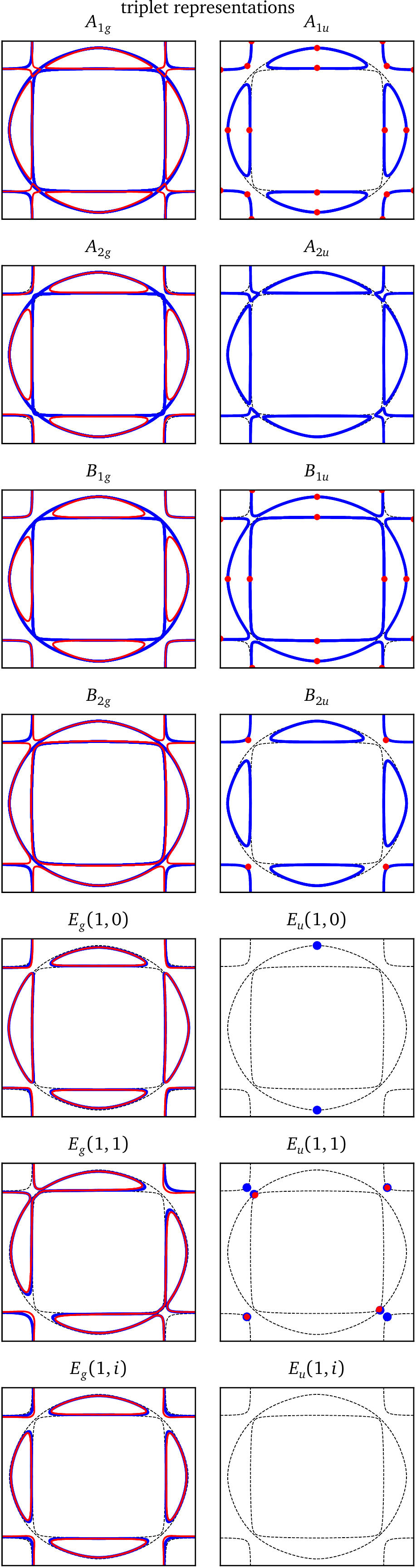}~~~
	\includegraphics[scale=0.6]{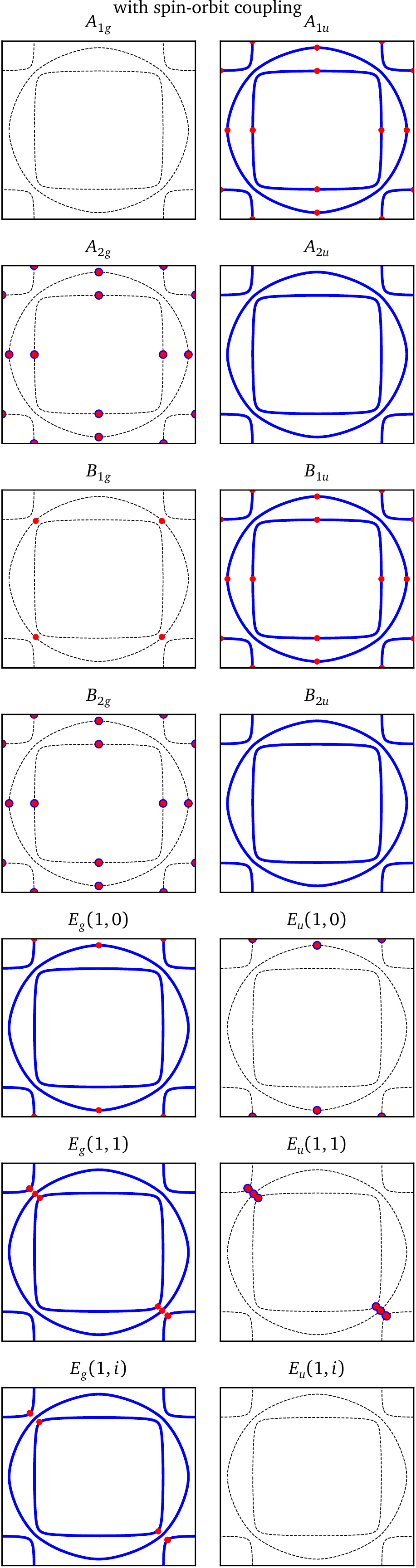}
	\caption{(color online) averaged or typical nodes associated to 
	the different irreps of $D_{4h}$ for $k_z=0$ (blue) and $k_z=\pi/2$ (red). Each panel covers the full Brillouin zone from $(-\pi,-\pi)$ to $(\pi,\pi)$ and the representation label is indicated on top.
	Left: singlet representations; middle: triplet representations; right: with spin-orbit coupling
	The normal state Fermi surface is the black dotted line.
	See text for details.
	}
	\label{fig:nodes}
	\end{figure*}

Column 4 of Table~\ref{table:gapS} shows the nodes associated with each function. The meaning of the symbols used is the following:
each of $\a$, $\b$ and $\g$ refers to the normal state Fermi surface sheets (see Fig.~\ref{fig:nodes_descr}) and when appearing alone, means that the whole sheet is a nodal surface. Nodal surfaces can be hybridized: For instance, a combination of the $\b$ and $\g$ surfaces, noted $(\b\g)$, is visible in the $B_{1u}$ panel of Fig.~\ref{fig:nodes}.
The $(\a\g)$ hybridization is seen in the $A_{1u}$ panel of the same figure, and a complete hybridization ($\a\b\g$) in the $E_g(1,1)$ panel.
When a Fermi surface sheet appears in conjunction with $+$, then the intersection of that sheet with horizontal and vertical axes at 0 and $\pm\pi$ constitute the nodes. The symbols $-$ and $|$ stand for horizontal and vertical lines only.
When appearing in conjunction with $\times$, then the intersection of that sheet with diagonals constitutes the nodes. The symbol $\diagdown$ stands for the north-west diagonal only. Commas separate different nodal lines or surfaces present. For two-dimensional representations ($E_g$ and $E_u$), we show the nodes obtained from the $(1,0)$, the $(1,1)$ and the $(1,i)$ combinations, separated by a colon.

\begin{figure}
	\includegraphics[scale=0.6]{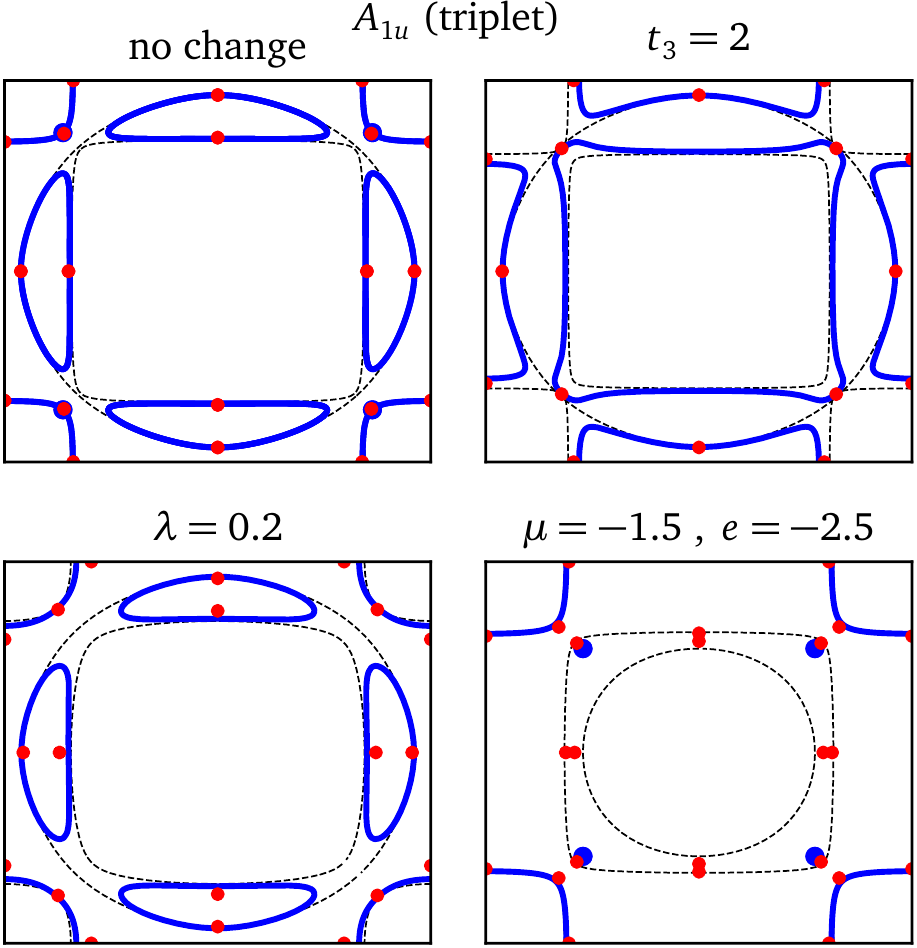}
	\caption{(color online) Averaged or typical nodes associated to the triplet representation $A_{1u}$, but with different band parameters.
	Top left: No change in the parameters. Other panels are obtained by changing the parameters as indicated on top of each panel.
	In particular, the bottom right panel corresponds to a drastic change in band parameters incompatible with \SRO.
	Again, the $k_z=0$ nodes are in blue (including equatorial nodes), and the $k_z=\pi/2$ vertical nodal lines are shown as red dots.
	The normal state Fermi surface is the black dotted line.
	}
	\label{fig:varia}
	\end{figure}

\subsection{On the notion of node imposed by symmetry}

Column 2 of Table~\ref{table:gapS} shows the nodes obtained when combining the different pairing functions of a given representation, with an equal amplitude of 0.25. Thus, this represents the approximate notion of ``nodes imposed by symmetry'' on each representation. These are in turn illustrated on the two leftmost columns of Fig.~\ref{fig:nodes}.
As a rule, the nodes of the combined pairing function in an irrep are the intersection of the nodes of the separate basis functions. The latter may separately have accidental nodes, but those generally disappear when taking linear combinations.

However, strictly speaking, \textit{the notion of symmetry-imposed nodes does not make sense in the case of multi-orbital models}, with or without spin-orbit coupling. In the one-band case, whose symmetry classification appears on Table~\ref{table:D4h}, a symmetry-imposed node corresponds to a pairing function that vanishes in some direction because it is odd under certain symmetry operations in that irrep. 
For instance, the pairing function must be odd under a diagonal reflexion $\s_d$ in the representation $B_{1g}$, and must accordingly vanish along the diagonals, which is indeed the case of the standard $d$-wave function $x^2-y^2$. 
The pairing function being a scalar, its zeros correspond to nodes.
Essentially, the one-band case is simple because translation invariance allows us to express the order parameter as a scalar function of the wavevector $\mathbf{k}$.

In a multi-orbital model, the pairing function is a multi-component objet: a matrix. That matrix may be odd under a certain symmetry operation, but that does not imply that it must vanish at a fixed point of that operation in momentum space, because the odd character can reside in the orbital part instead of the spatial part. Indeed, the odd character translates into the following transformation property for the pairing function:
\begin{equation}
\Delta_{\nu}(x,y,z) \to \Delta'_{\nu}(x,y,z) = \mathcal{U}(\s_d)_{\nu\nu'}\Delta_{\nu'}(y,x,z) 
\end{equation}
where the index $\nu$ labels basis vectors in orbital space and $\mathcal{U}$ the orbital part of the representation. 
In the $B_{1g}$ representation, we therefore have the condition
$\Delta'_{\nu}(x,y,z) = -\Delta_{\nu}(x,y,z)$, or $\mathcal{U}(\s_d)\Delta(y,x,z) = -\Delta_{\nu}(x,y,z)$,
which translates into $\mathcal{U}(\s_d)\Delta(x,x,z) = -\Delta_{\nu}(x,x,z)$ on the diagonal.
In the single-orbital case, $\mathcal{U}=1$ and that condition implies $\Delta(x,x,z)=0$. In the multi-orbital case, the pairing function may be an eigenvector of 
$\mathcal{U}$ with eigenvalue $-1$, and this imposes no condition at all on $\Delta(x,x,z)$.
For instance, the pairing function $\ax-\ay$, which is wavevector independent, belongs to $B_{1g}$. The matrix $\mathcal{U}$ in that case
exchanges $a_x$ and $a_y$ and is equivalent to $-1$ in orbital space, which leaves an even (here constant) spatial part.

As another example, the inter-orbital pairing function $\cx x + \cy y$ in representation $A_{1u}$ describes a singlet state that is odd under the reflexion $\s_z$ with respect to the $xy$-plane.
Indeed, under this reflexion, the orbitals $d_{xz}$ and $d_{yz}$ change sign, and so do the components $c_x$ and $c_y$, while the functions $x$ and $y$ are unaffected.
The matrix-valued pairing function then takes the form
\begin{equation}
\Delta(x,y,z) =
\begin{pmatrix} 0 & 0 & y \\ 0 & 0 & x \\ -y & -x & 0 \end{pmatrix}
\end{equation}
(we ignore spin, which is in a singlet state in this example). The transformation law of that pairing function under $\s_z$ is $\Delta\to\Delta'=U(\s_z)\Delta U(\s_z)$, where $U(\s_z)$ is given in Table~\ref{table:generators}. Therefore $\Delta'=-\Delta$, as it should be in representation $A_{1u}$.
Accordingly, while that pairing function has nodes (in fact nodal surfaces, since nothing depends on $z$ here), their precise shape is not imposed by symmetry. In particular they do not coincide with the normal Fermi surfaces, but are rather hybridized Fermi surfaces, as illustrated on Fig.~\ref{fig:nodes}.

Some of the combined nodes illustrated in Fig.~\ref{fig:nodes} are therefore generic in their character (point or surface) but accidental in their precise shape. Depending on the precise coefficients of the combined pairing function, the precise shape of a hybridized nodal surface may vary slightly.

Table \ref{table:gapT} lists the triplet pairing functions found using the same procedure. In that case only products of orbital and spatial functions that are antisymmetric under electron exchange were kept.
The combined nodes are illustrated on the middle two columns of Fig.~\ref{fig:nodes}.

In this figure, we have shown the nodes found on the $k_z=0$ plane (in blue) and those on the $k_z=\pi/2$ plane (in red). The blue curves on the figure thus correspond to horizontal (more precisely, equatorial) nodal lines. 
A majority of representations have them. Often nodal lines also occur at $k_z=\pi/2$ but in a hybridized form, hinting at a complex three-dimensional representation of the nodes in those cases. 
Note that our tight-binding model is still strictly two-dimensional.
In no case do the generic nodes coincide with the normal state Fermi surfaces. In that sense, superconductivity is never \textit{hidden} in this system, even though it can in many cases be called \textit{gapless}, since the nodes occur at every angle, at least in the absence of spin-orbit coupling.

An important point is that the only two representations that have no nodes are the singlet $A_{1g}$, which we could commonly call $s$-wave, and the triplet $E_u(1,i)$, which we could call $p_x+ip_y$.
This is still true with spin-orbit coupling.

In order to illustrate how these nodes vary upon changing the band parameters, we have plotted the typical nodes for three additional sets of band parameters on Fig.~\ref{fig:varia}.
The details of the nodal surfaces change, but the presence of nodal lines along various axes is robust.

\subsection{Spin orbit coupling}

In the presence of the spin-orbit coupling ($\kappa\ne0$), the symmetry is reduced. The spin will transform according to the generators listed in table~\ref{table:generators}, within a spin representation of $D_{4h}$, not listed in the character table~\ref{table:D4h}. In particular, within such a spin representation, the fourth power $S(C_4)^4$ is $-1$, not 1.
The tensor product of this spin representation with itself yields symmetric and antisymmetric representations, characterized by the $\dv$-vector components. These in turn can be tensored with orbital and spatial representations, provided the overall pairing function is antisymmetric.

Table \ref{table:gapSO} lists the possible pairing functions in the presence of spin-orbit coupling. The format used is the same as in Tables \ref{table:gapS} and \ref{table:gapT}.
Note, however, that the Fermi surface of the normal state (the dotted line) is slightly different, because of the added spin-orbit term $\kappa$.

The generic nodes of a given representation in the spin-orbit case are generally the intersections of the nodes of the corresponding singlet and triplet representations, although this is not always the case, maybe because the spin-orbit coupling changes the normal-state dispersion as well.
Overall, the situation is a bit simpler with spin-orbit coupling: 3D nodal surfaces dot not exist: only equatorial and vertical nodal lines do.
Half of the representations have equatorial nodes. The only representation without nodes are $A_{1g}$ (or $s$-wave) and $E_u(1,i)$ (or $p_x+ip_y$).

\begin{table}
	\caption{
	Character table of $D_{2h}$.\label{table:D2h}
	}
	\begin{ruledtabular}
	\begin{tabular}{LRRRRRRRR} 
	& E & C_{2z} & C_{2y} & C_{2x} & i & \sigma_z & \sigma_y & \sigma_x \\[4pt] \hline\\[-6pt]
	A_{g} & 1 & 1 & 1 & 1 & 1 & 1 & 1 & 1 \\ 
	B_{1g} & 1 & 1 & -1 & -1 & 1 & 1 & -1 & -1 \\
	B_{2g} & 1 & -1 & 1 & -1 & 1 & -1 & 1 & -1 \\
	B_{3g} & 1 & -1 & -1 & 1 & 1 & -1 & -1 & 1 \\ 
	A_{1u} & 1 & 1 & 1 & 1 & -1 & -1 & -1 & -1 \\
	B_{1u} & 1 & 1 & -1 & -1 & -1 & -1 & 1 & 1 \\
	B_{2u} & 1 & -1 & 1 & -1 & -1 & 1 & -1 & 1 \\
	B_{3u} & 1 & -1 & -1 & 1 & -1 & 1 & 1 & -1
		\end{tabular}
	\end{ruledtabular}
	\end{table}
	
\begin{figure}
	\includegraphics[width=\hsize]{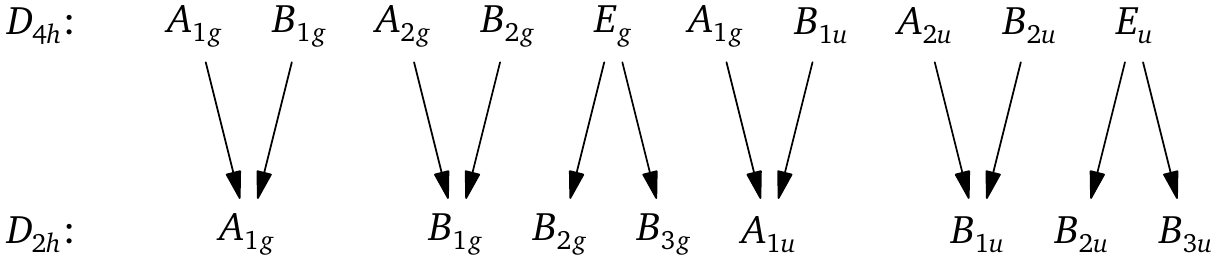}
	\caption{Branching of the irreducible representations of $D_{4h}$ (top) into those of $D_{2h}$ (bottom).
	}
	\label{fig:branching}
	\end{figure}

\begin{figure*}
\includegraphics[scale=0.6]{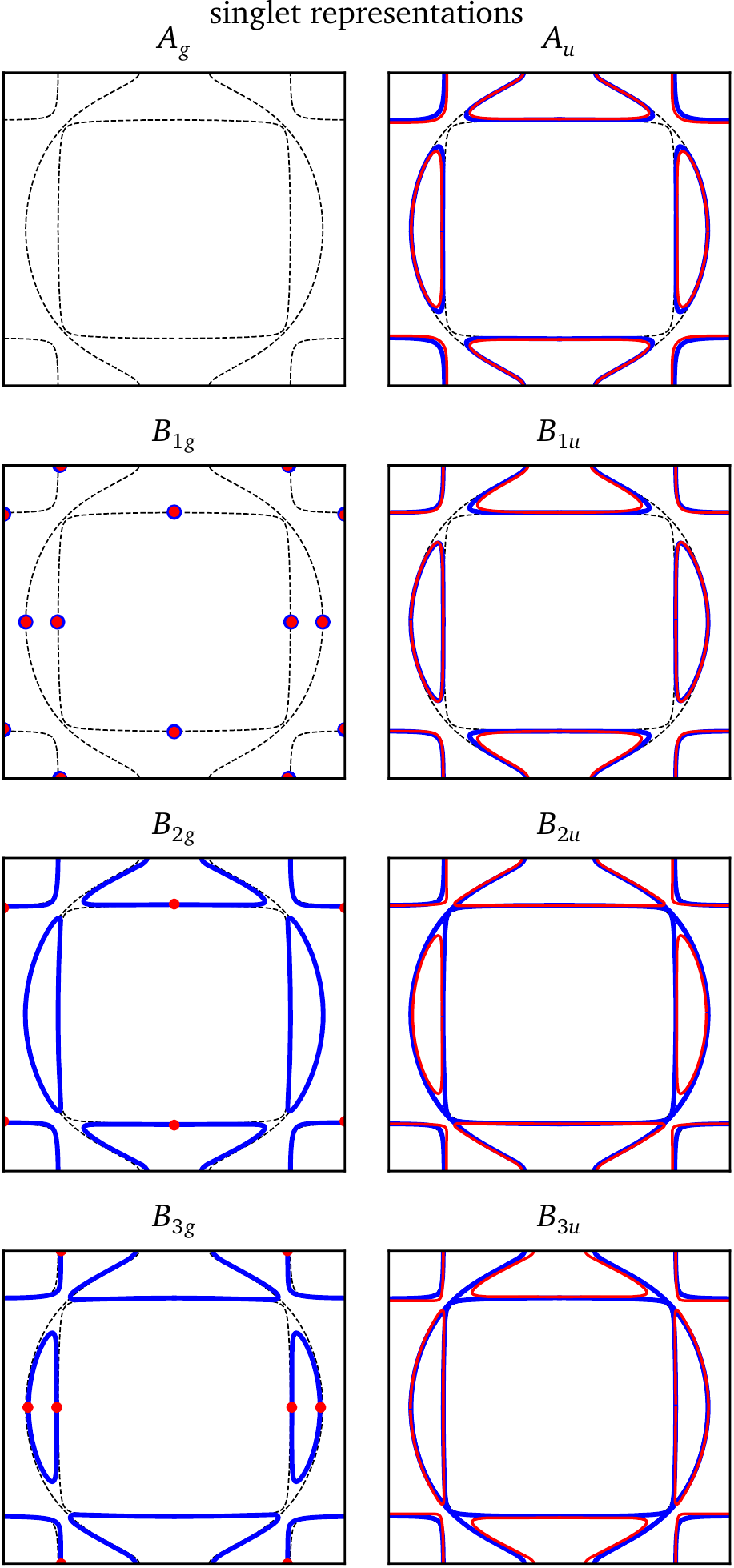}~~~
\includegraphics[scale=0.6]{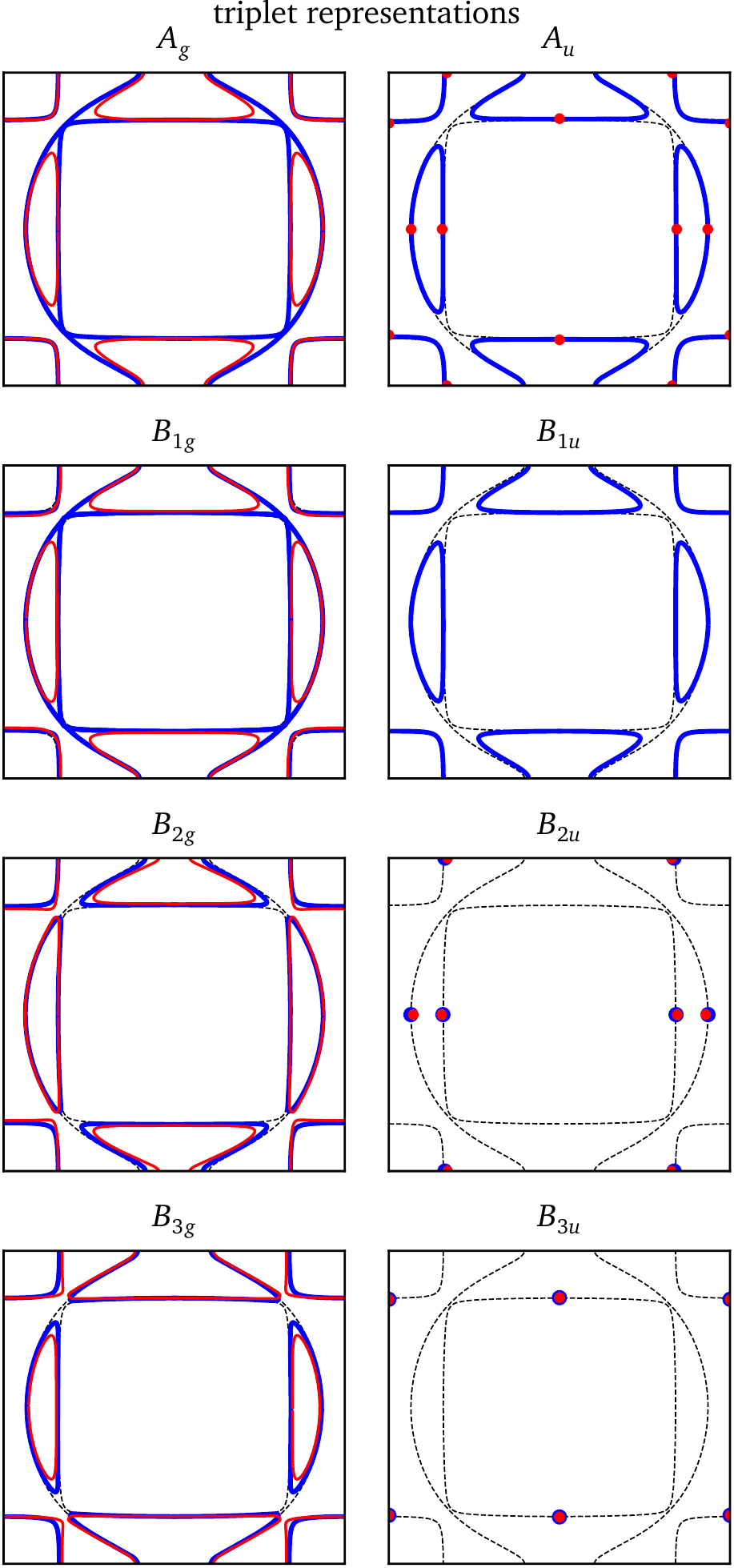}~~~
\includegraphics[scale=0.6]{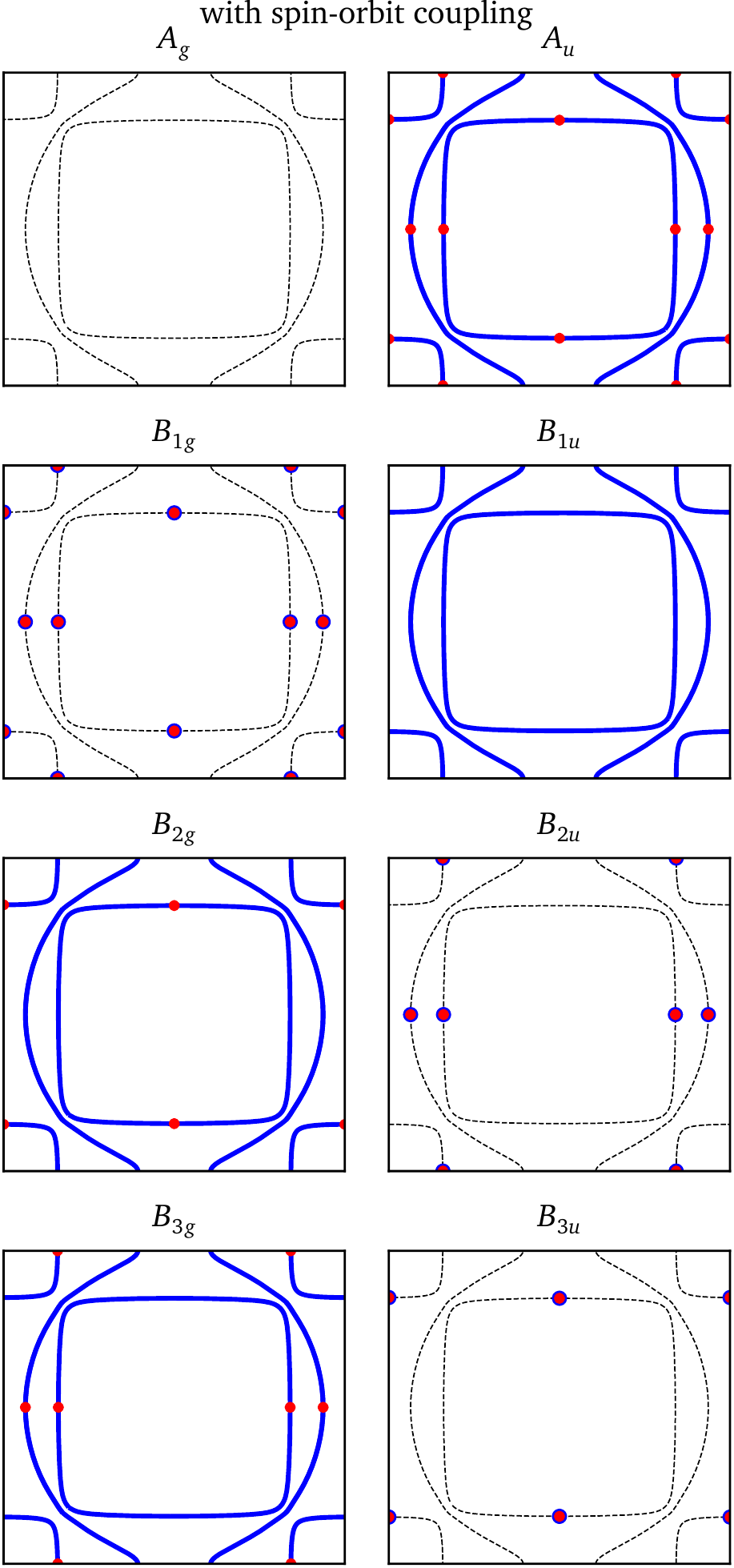}
\caption{(color online) averaged or typical nodes associated to 
the different irreducible representations of $D_{2h}$ for $k_z=0$ (blue) and $k_z=\pi/2$ (red). Each panel covers the full Brillouin zone from $(-\pi,-\pi)$ to $(\pi,\pi)$ and the representation label is indicated on top.
Left: singlet representations; middle: triplet representations; right: with spin-orbit coupling
The normal state Fermi surface is the black dotted line.
}
\label{fig:nodes_D2h}
\end{figure*}

\subsection{Uniaxial deformations}\label{sec:D2h}

Under uniaxial pressure along the $x$ or $y$ axis, \SRO\ will undergo a slight spatial deformation that will reduce its point-group symmetry from $D_{4h}$ to $D_{2h}$. In this subsection, we outline the changes that this would bring to the classification explained above.

$D_{2h}$ is an Abelian subgroup of $D_{4h}$ and contains half of its elements.
Basically, the generator $C_4$ is no longer a symmetry operation and all group elements obtained from it drop out. The character table of $D_{2h}$ is reproduced in Table~\ref{table:D2h}. The irreps of $D_{4h}$ collapse into the irreps of $D_{2h}$, as illustrated schematically in Fig.~\ref{fig:branching}.

A similar analysis as done above can be carried out for the $D_{2h}$ symmetry, after adding an anisotropy parameter $\a=0.05$ such that hopping parameters $t_1$ and $t_3$ are augmented by $\a$ in the $x$ direction and diminished by the same amount in the $y$-direction. This small value of $\a$ is sufficient to make the $\g$-band Fermi surface open, i.e., to bring about a Lifshitz transition, as observed in experiments~\cite{Sunko2019}.
The resulting nodes are illustrated on Fig.~\ref{fig:nodes_D2h}. The main change from the isotropic case is the disappearance of chiral representations. Thus, the breaking of time-reversal symmetry could only occur by combining different irreps.
In particular, the only representation that has no nodes at all is $A_{1g}$ ($s$-wave).

\section{Discussion and conclusion}\label{sec:discussion}

The main novelty introduced in this paper is the integration of inter-orbital pairings, in particular odd-orbital pairings, in the classification of superconductivity for $t_{2g}$ systems. 
However, at this point we are not in a position to say that this kind of superconductivity is present in \SRO. 
Indeed, in addition to suggestions for inter-orbital pairing~\cite{puetter2012,olivier}, there are also good arguments to indicate that these kind of pairings should not be favored in the weak-coupling limit~\cite{ramires2016}. 
We can still highlight the main differences between inter-orbital and single-orbital superconductivity, and see how they constrain the interpretation of available experimental data for \SRO.

\subsection{Singlet vs triplet superconductivity}

An odd orbital part for the superconducting order parameter (i.e., involving the $\cv$ vector) allows the combinaison of singlet and odd parity, or triplet and even parity order parameters. This contrasts with the single-orbital case where singlet and triplet respectively imply even and odd parity. 
Here, both singlet and triplet can be associated with any $e$-type or $u$-type representation.
In the presence of spin-orbit coupling, this implies that there is no clear distinction between singlet and triplet and that a combinaison of both is possible in general, as seen in table~\ref{table:gapSO}.
Some studies \cite{veenstra2014, zhang2016} have suggested the possibility of combining singlet and triplet order parameters in \SRO\ due to strong spin-orbit coupling. 
Our analysis shows that the only way to achieve such combinaisons whithin the same irreducible representation is through odd orbital pairing.

\subsection{Odd vs even superconductivity}

As can be seen from Table~\ref{table:generators}, the inversion operation $i=\s_x\s_y\s_z$ has no effect on the orbitals, and therefore on the $\av$, $\bv$ and $\cv$ vectors. This implies that all states within an irrep have the same spatial parity. In particular, all $g$-type representations are even and all $u$-type representations are odd.
Josephson interferometry experiments~\cite{Nelson1151, laube2000, liu2010} have suggested that the order parameter of \SRO\ has odd parity.
If this were true, it would eliminate all the $g$-type representations

\subsection{Broken time-reversal symmetry}

Broken time-reversal symmetry, supported by muon spin resonance \cite{btrs} and polar Kerr effect \cite{kerr} experiments, can only occur in our paradigm within two-dimensional irreps ($E_g$ and $E_u$). However, the absence of splitting of the superconducting transition when applying strain seems to exclude that possibility. We are facing a contradiction that cannot be resolved without abandoning the single phase transition hypothesis; the possibility of inter-orbital pairing is of no help here.


\subsection{Nodes and the density of states}

One of the main motivations of this work was to predict typical nodal structures from symmetry considerations.
We have seen that the notion of nodes imposed by symmetry is not strictly valid when many orbitals are involved in the superconducting state.
However, there are typical nodes that can be observed in a given irreducible representation, and they are shown on Fig.~\ref{fig:nodes}.
There is contradicting evidence for both vertical nodal lines \cite{specific-heat,taillefer,lupien2001} and horizontal nodal lines \cite{horizontal-nodes} in \SRO. 

However, nodal \textit{surfaces} would lead to a finite density of states at the Fermi level within the superconducting state, which seems excluded~\cite{taillefer}.
Simple single-orbital pairing functions involving only $\az$, or a combination of $\ax$ and $\ay$, would lead to nodal surfaces coinciding with the Fermi surfaces of the bands not involved in pairing. It is likely, however, that interactions would cause superconductivity to have components in every band. Fig.~\ref{fig:nodes} shows that $u$-type singlet representations and $g$-type triplet representations have nodal surfaces. These disappear when spin-orbit coupling is important.

If we exclude two-dimensional representations, keeping nodal vertical and horizontal nodal lines would tend to favor representations $A_{1u}$ and $B_{1u}$ if spin-orbit coupling is important. Incidently, both of these are odd under inversion, which is also supported by observations~\cite{Nelson1151}.

From a strongly interacting perspective, it makes sense to seek real-space pairing along the same bonds as the most important hopping terms. Therefore we are led to favor the lowest possible degree in pairing functions, as they correspond to the shortest ranges, and to exclude pairing functions in the $z$ direction. It is, however, difficult to meet this requirement while considering a gap with horizontal nodes, with or without inter-orbital pairing.
For instance, the spin-orbit irreps $A_{1u}$ and $B_{2u}$ have horizontal and vertical nodes, no nodal surfaces; but the simplest pairing functions belonging to these representations
(from Table~\ref{table:gapSO}) involve  inter-orbital, nearest-neighbor pairing, which does not correspond to hopping terms of the model studied.
On the other hand, the spin-orbit $E_g$ representations also have the correct nodal content, have constant inter-orbital pairing functions, and even allow for a broken time-reversal solution ($E_g(1,i)$). Furthermore, these nodes are preserved even as uniaxial pressure is applied (see Fig.~\ref{fig:nodes_D2h} under $B_{2g}$ and $B_{3g}$). However, as mentioned above, the absence of transition splitting when applying uniaxial pressure does not favor representations of this type, and they are not odd under inversion.

Finally, let us remark that our results are can easily be applied to other $t_{2g}$ systems with $D_{4h}$ symmetry. The precise values of the hopping terms are not important in the classification we presented, although some fine details about the shape of the nodes will vary, as illustrated on Fig.~\ref{fig:varia}.

\begin{acknowledgments}
Fruitful discussions with J.~Clepkens, O.~Gingras, R.~Nourafkan and A.-M.~Tremblay are gratefully acknowledged.
This work has been supported by the Natural Sciences and Engineering Research Council of Canada (NSERC) under grant RGPIN-2015-05598 and the Canada First Research Excellence Fund.
\end{acknowledgments}


%
	
\end{document}